\documentclass[conference]{IEEEtran}
\IEEEoverridecommandlockouts
\usepackage{cite}
\usepackage{amsmath,amssymb,amsfonts}
\usepackage{algorithmic}
\usepackage{graphicx}
\usepackage{textcomp}
\usepackage{xcolor}
\usepackage{subcaption}
\def\BibTeX{{\rm B\kern-.05em{\sc i\kern-.025em b}\kern-.08em
    T\kern-.1667em\lower.7ex\hbox{E}\kern-.125emX}}
\begin{document}

\title{COMPARATIVE ANALYSIS OF SWITCHING DYNAMICS IN DIFFERENT MEMRISTOR MODELS}

\author{\IEEEauthorblockN{1\textsuperscript{st} Santosh Parajuli}
\IEEEauthorblockA{\textit{Department of Electrical and Electronics Engineering} \\
\textit{Kathmandu University}\\
Kathmandu, Nepal \\
eesantosh@ku.edu.np}
\and
\IEEEauthorblockN{2\textsuperscript{nd} Ram Kaji Budhathoki}
\IEEEauthorblockA{\textit{Department of Electrical and Electronics Engineering} \\
\textit{Kathmandu University}\\
Kathmandu, Nepal \\
ram.budhathoki@ku.edu.np}

}

\maketitle

\begin{abstract}
Memristor, memory resistor, is an emerging technology for computational memory. Number of different memristor models are available based on the physical experiments. 
To use memristor as a computational memory element, one should know how the internal state modulates in time when driven by current or voltage. In this paper, we examine three widely used models and  make a comparison of how internal state in these models changes with respect to input current or voltage. In Strukov model, internal state changes linearly with the input current. However, the linearity of internal state modulation in Yang model can be controlled. On the other hand, Pickett model shows non linear variation in internal state with the input current.    
 
\end{abstract}


\section{Introduction}
Memristor was first postulated by Leon Chua in 1971 as fundamental circuit element \cite{chuamem}. In a memristor, when current flows in  one direction, its resistance decreases and vice versa. Hence, Memristor changes its state with the input current. Present state  of a memristor is the cumulative effect of previous currents. The state is hold at a particular voltage or resistance level if the current flow is suddenly stopped. Therefore, looking at the present memristor state, past current through the device can be computed.

From a circuit perspective, memristor is often modeled as a two terminal device which has top and bottom electrodes and sandwiched conductive channel, which acts as a switching layer \cite{stanley}-\cite{yang}. 
The definition of memristive stystems \cite{kang} is given in general by the equations 
\begin{equation}
\label{Eq: 1}
\frac{dx}{dt}= f(x,u,t)
\end{equation}
\begin{equation}
\label{Eq: 2}
y=g(x,u,t)u
\end{equation}
where $u$ and $y$ denote the input and output of the system, and $x$ denotes the state of the system. From \eqref{Eq: 1}, in a memristive system, the state evolves with time $t$. This state modulation in a memristive system closely relates with strengthening or weakening of synapses over time in a neuromorphic system\cite{eric}. The equations can be modified by choosing appropriate variables to define memristor for electronic circuit as
\begin{equation}
\label{Eq: 1modified}
\frac{dw}{dt} = f(w,v_M,t)
\end{equation}
\begin{equation}
\label{Eq: 2modified}
i_M=M(w,v_M,t)v_M
\end{equation}
where $v_M$ and $i_M$ denote the current and voltage of the memristor, and $w$ denotes the internal state variable. $M$ denotes the memconductance (memristance). As shown in \eqref{Eq: 1modified}, the voltage input directly affects the state variable $w$, which is typical in first order memristive system \cite{yeonjoo}.

Different physical models are proposed for the memristive system  \cite{stanley}-\cite{yang},  \cite{Shahar}-\cite{fernando}. In this paper, we first 
examine memristor models proposed by Strukov \textit{et al.} \cite{stanley}, Yang \textit{et al.} \cite{yang}, and Pickett \textit{et al.} \cite{pickett}, and then do comparative analysis of how internal state varies with respect to time in these models.

\section{Memristor Models}
In this work, we take three memristor models derived from the result of physical experiment at the nanoscale level. All three models meet the definition of memristive system given in \eqref{Eq: 1} and \eqref{Eq: 2}. 
\subsection{Linear Memristor Model Proposed by Sturkov \textit{et al.}}
\begin{figure}[h]
\center
\includegraphics[scale=0.6]{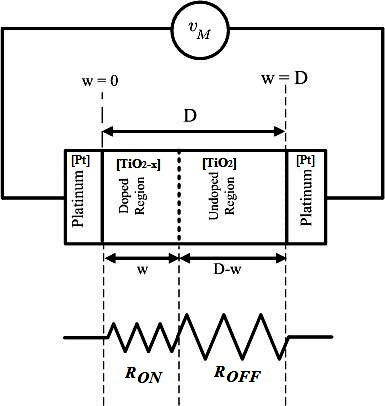}
\caption{Equivalent circuit for Strukov model: Two state dependent resistors in series.}
\label{Fig: Strukov}
\end{figure}

This model shows how memristance arises naturally in nanoscale system due to coupling of electronic and ionic transport under an external bias. Equivalent circuit for this model is shown in Fig. \ref{Fig: Strukov}.
The device dynamics and the I-V characteristics are described by the equations 
\begin{equation}
\label{Eq: 4}
\frac{dw}{dt} = \mu_v \frac{R_{ON}}{D}i_M
\end{equation}
\begin{equation}
\label{Eq: 5}
v_M = \left(R_{ON}\frac{w}{D}+R_{OFF}\left(1-\frac{w}{D}\right)\right )i_M
\end{equation}
where $\mu_v$ is average ion mobility, $R_{ON}$ is the resistance of doped region, $R_{OFF}$ is the resistance of undoped region,  and $w \in (0,D)$. 

This model (a) considers ohmic electronic conduction and linear ionic drift (b) assumes modulation of $w$ is due to drift of charged dopants (c) does not reflect change in the interface ($Pt/TiO_2$) due to switching. (d) does not consider tunneling effect.
\begin{figure}[t]
\center
\includegraphics[width=\linewidth]{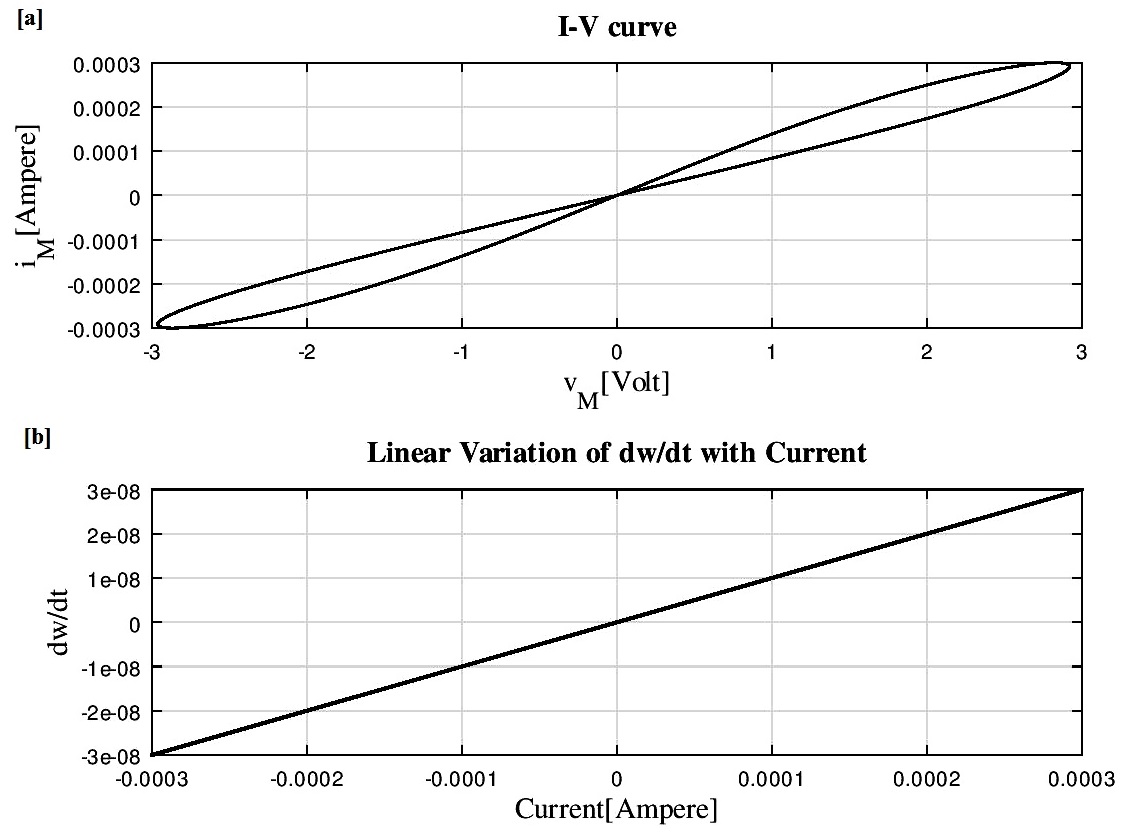}
\caption{Simulation result of Strukov \textit{et al.} model with $R_{ON}=100 \Omega$, $R_{OFF}=16 k\Omega$, $D= 10 nm$, $frequency= 3 Hz$, driven by a sinusoidal source having zero average without taking the memristor into saturation.}
\label{Fig: Strukovsimulation1}
\end{figure}

\subsection{Non-Linear Memristor Model Proposed by Yang \textit{et al.}}

\begin{figure}[h]
\center
\includegraphics[width=\linewidth]{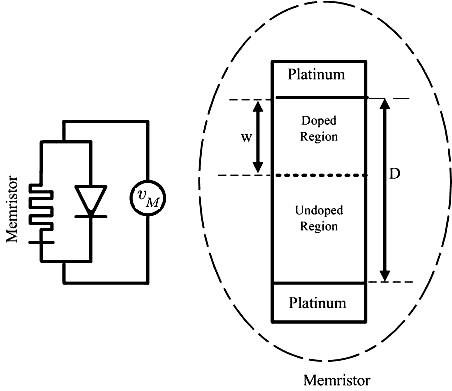}
\caption{Equivalent circuit for Yang \textit{et al.} model: A diode in parallel with a memristor (shown in dotted circle).}
\label{Fig: Yang}
\end{figure}

\begin{figure}[t]
\center
\includegraphics[width=\linewidth]{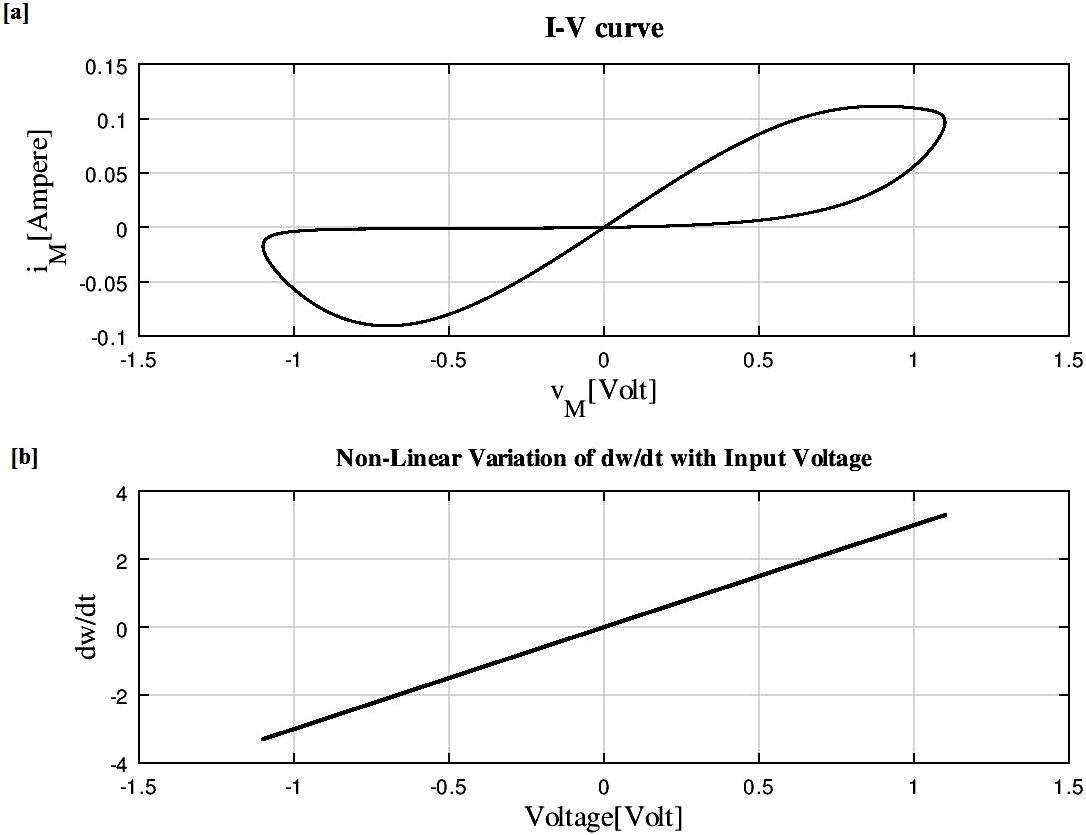}
\caption{Simulation result of Yang \textit{et al.} model with $R_{ON}=100 \Omega$, $R_{OFF}=16 k\Omega$, $D= 10 nm$, $m=1$, $frequency= 3 Hz$, driven by a sinusoidal source having zero average without taking the memristor into saturation.}
\label{Fig: yangsimulation1}
\end{figure}

Strukov \textit{et al.} model neglects the important role of Pt-TiO$_2$ interface; however, this model considers change to the electronic barrier at the interface due to switching. Based on the physical experiments, Strukov \textit{et al.} demonstrate that oxygen vacancy drift towards the interface to create conductive channels, and away from the interface to eliminate such channels. Equivalent circuit for this model is shown in Fig. \ref{Fig: Yang}, which consists a rectifier in parallel with a memristor. 
The device dynamics \cite {cnn} and the I-V characteristics \cite{yang} are described by the equations 

\begin{equation}
\label{Eq: 7}
\frac{dw}{dt} = \alpha v^m_M
\end{equation} 

\begin{equation}
\label{Eq: 8}
i_M = w^n\beta sinh(\delta v_M)+\chi \left[ exp(\gamma v_M)-1 \right]
\end{equation}
where $\alpha$ is a constant and $m$ is an odd integer. If $m=1$ in \eqref{Eq: 7}, this model reduces to  Strukov \textit{et al.} model. $\beta$, $\delta$, $\chi$, $\gamma$ are experimental fitting parameters, and $n$ determines the influence of the state variable on the current.
\begin{figure}[b]
\center
\includegraphics[width=\linewidth]{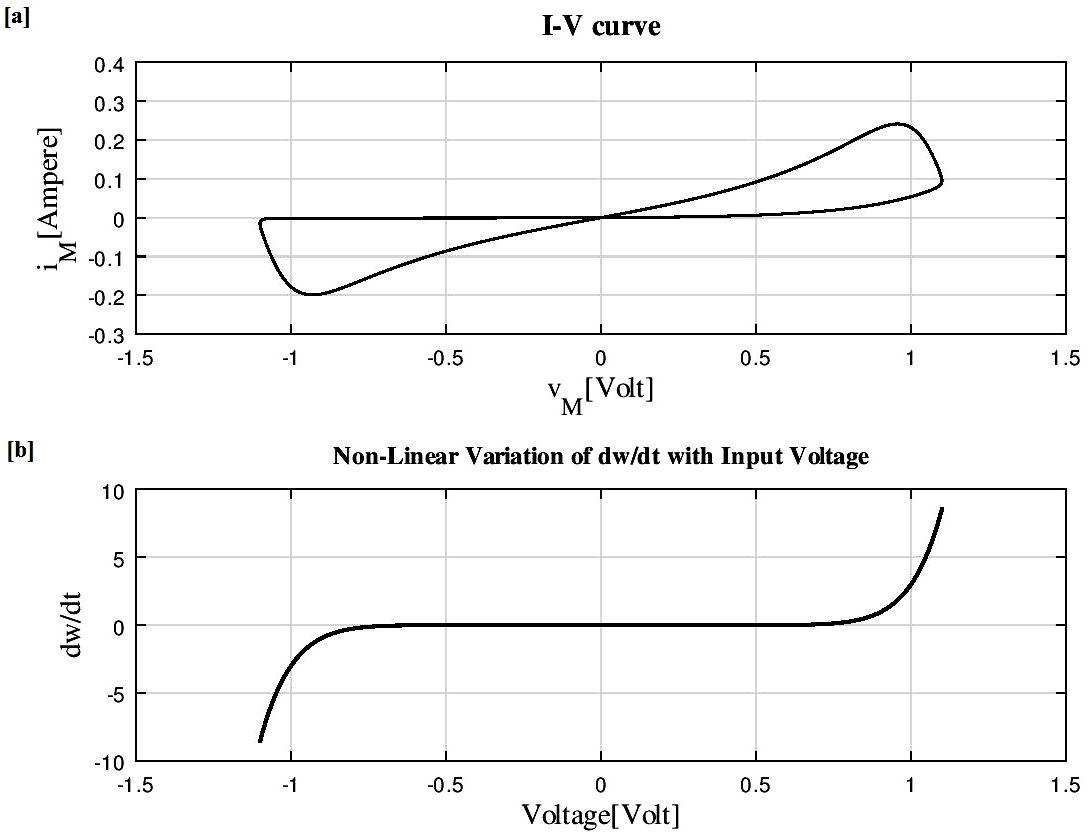}
\caption{Simulation result of Yang \textit{et al.} model with $R_{ON}=100 \Omega$, $R_{OFF}=16 k\Omega$, $D= 10 nm$, $m=11$, $frequency= 3 Hz$, driven by a                 sinusoidal source having zero average without taking the memristor into saturation.}
\label{Fig: yangsimulation11}
\end{figure}
I-V and dw/dt vs. Volt characteristics revel the existence of threshold voltage around 0.7 for the applied voltage of 2.2 peak to peak amplitude. 
The reduction of Yang \textit{et al.} model into Strukov \textit{et al.} model with $m=1$ is shown in Fig. \ref{Fig: yangsimulation1}.    

\subsection{Non-Linear Memristor Model Proposed by Pickett \textit{et al.}}
\begin{figure}[b]
\center
\includegraphics[width=\linewidth]{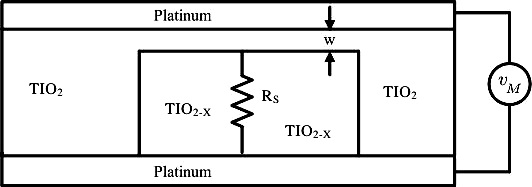}
\caption{Equivalent structural model proposed by Pickett \textit{et al.} which consists of an ohmic resistor in series with a tunneling barrier of width $w$.}
\label{Fig: 1}
\end{figure}

Based on the physical experiments, this model assumes that electronic conduction in Pt-TiO$_2$-Pt device is primarily due to modulation of effective tunneling barrier width under an applied voltage or current.   
By changing the tunneling gap, memristor can be switched between ON state, and OFF state; moreover, this dynamical behavior for off and on switching is highly nonlinear and asymmetric. Equivalent circuit for this model is shown in Fig. \ref{Fig: 1}. 
The device dynamics and I-V characteristics \cite{jsimmon}, \cite{yakopic}  are described by the equations\\

For OFF switching, $i_M>$0
\begin{equation}
\label{Eq: 9}
\frac{dw}{dt} = f_{\text{off}}\text{sinh}\left(\frac{i_M}{i_{\text{off}}} \right)exp\left[-exp\left(\frac{w-a_{\text{off}}}{w_c}-\frac{|i_M|}{b} \right)-\frac{w}{w_c} \right]
\end{equation}

For ON switching, $i_M<$0
\begin{equation}
\label{Eq: 10}
\frac{dw}{dt} = f_{\text{on}}\text{sinh}\left(\frac{i_M}{i_{\text{on}}} \right)exp\left[-exp\left(\frac{w-a_{\text{on}}}{w_c}-\frac{|i_M|}{b} \right)-\frac{w}{w_c} \right]
\end{equation}
\begin{equation}
\label{Eq: 11}
i_M = \frac{0.0617}{\Delta w^2}\left\lbrace
\phi_Ie^{-B\sqrt{\phi_I}}-(\phi_I+|v_g|)e^{-B\sqrt{\phi_I+|v_g|}}
\right\rbrace
\end{equation}
where $f_{off}$, $i_{off}$, $a_{off}$, $b$, $w_c$,$f_{on}$, $i_{on}$, $a_{on}$, are experimental fitting parameters, 
$$
\phi_I = \phi_0 - |v_g|\left(
\frac{w_1+w_2}{w(t)}
\right)-\left(
\frac{0.1148}{\Delta w}ln\left(
\frac{w_2(w(t)-w_1)}{w_1(w(t)-w_2)}
\right)
\right)
$$
$$
w_2 = w_1 + w(t)\left( 
1-\frac{9.2\lambda}{2.85+4\lambda-2|v_g|}
\right)
$$
$v_g$ represents voltage across $\text{TIO}_2$ layer.
$$
v_M  = v_g + \text{Drop across R}_\text{S}
$$
 $\phi_0 = 0.95 V$, $w_1 = 0.1261 nm$, $B=10.24634\Delta w$, $\Delta w = w_2-w_1$, $\lambda = \frac{0.0998}{w(t)}$.

\begin{figure}[t]
\center
\includegraphics[width=\linewidth]{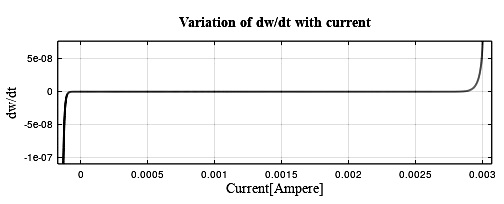}
\caption{Simulation result of Pickett \textit{et al.} model driven by a sinusoidal source having zero average without taking the memristor into saturation.The graph clearly shows higher positive and lower negative threshold currents.}
\label{Fig: pickettsimulation1}
\end{figure}
Unlike Strukov \textit{et al.} and Yang \textit{et al.} model, the state in this model changes asymmetrically with time, which is shown in Fig. \ref{Fig: pickettsimulation1}. The time taken for ON to OFF transition is different from OFF to ON transition.


\section{Memristor Switching Dynamics}
\begin{figure}[t]
\begin{subfigure}{.5\textwidth}
  \centering
  \includegraphics[width=.5\linewidth]{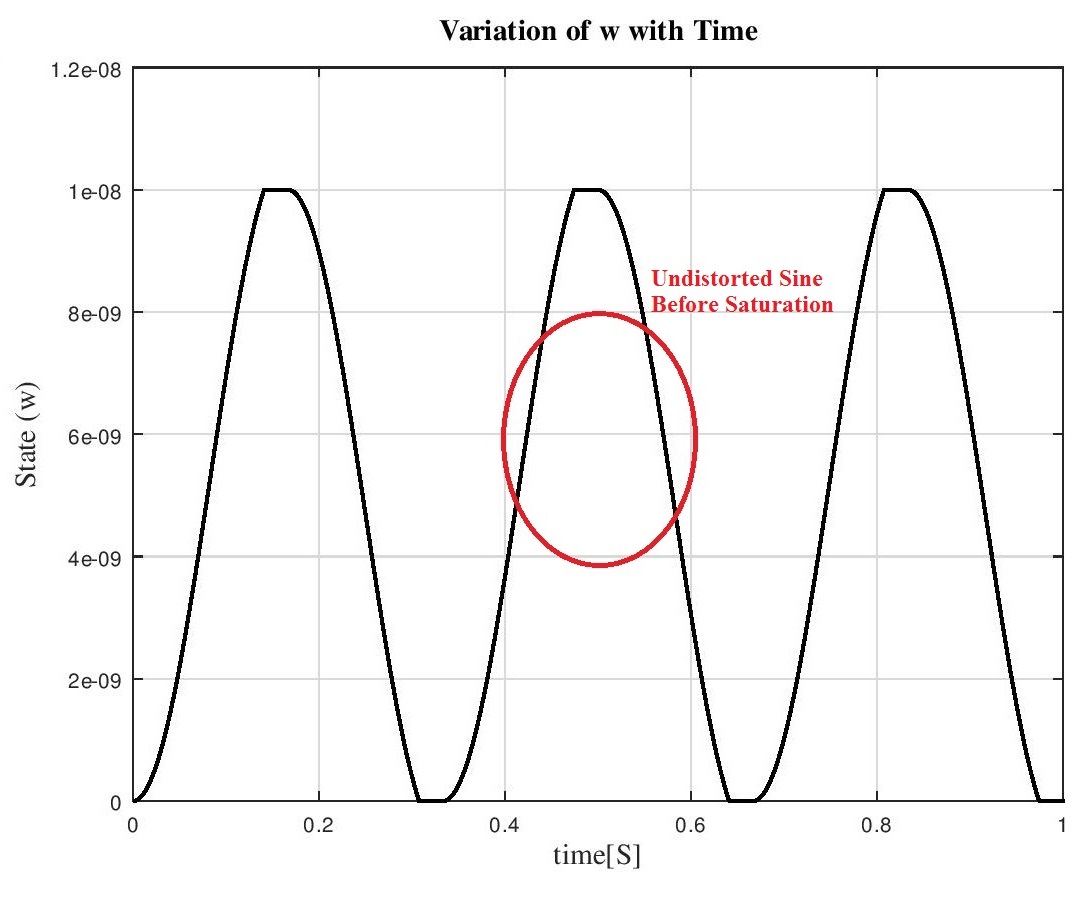}  
  \caption{Switching dynamics in Strukov \textit{et al.} model}
  \label{fig:6a}
\end{subfigure}
\begin{subfigure}{.5\textwidth}
  \centering
  \includegraphics[width=.5\linewidth]{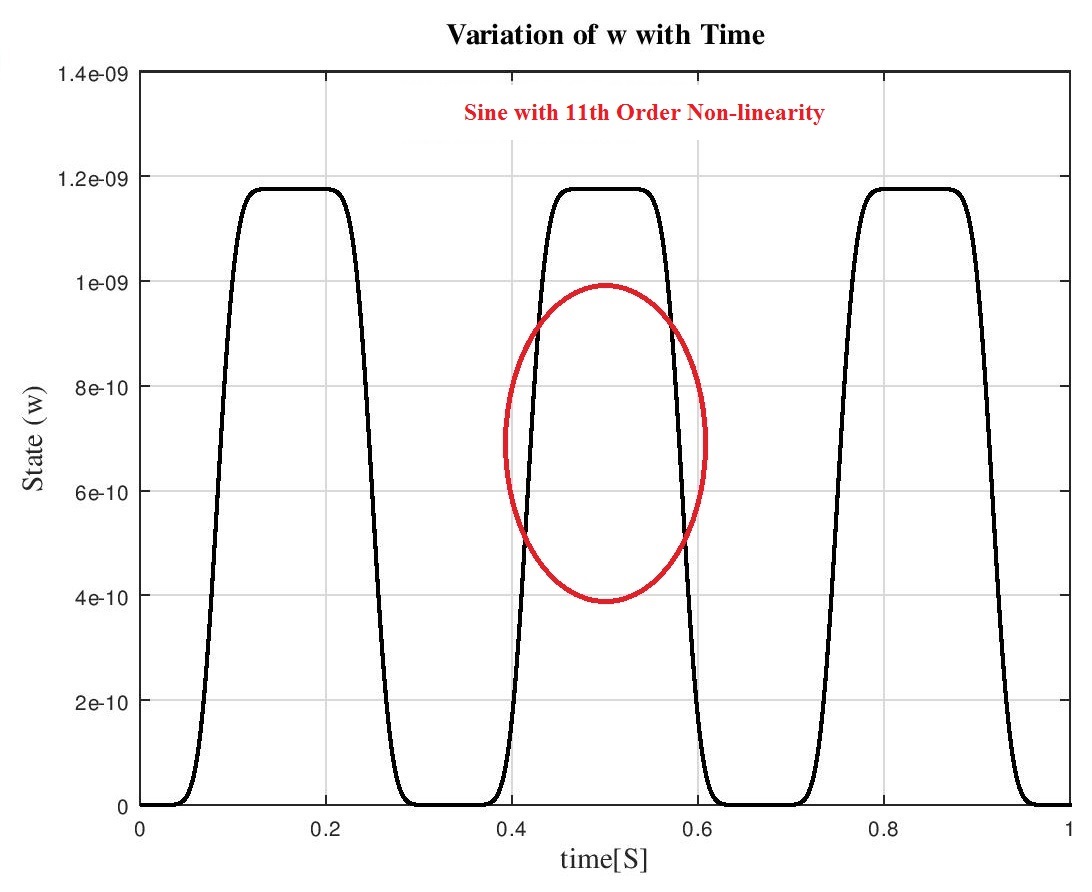}  
  \caption{Switching dynamics in Yang \textit{et al.} model}
  \label{fig:6b}
\end{subfigure}

\begin{subfigure}{.5\textwidth}
  \centering
  \includegraphics[width=.5\linewidth]{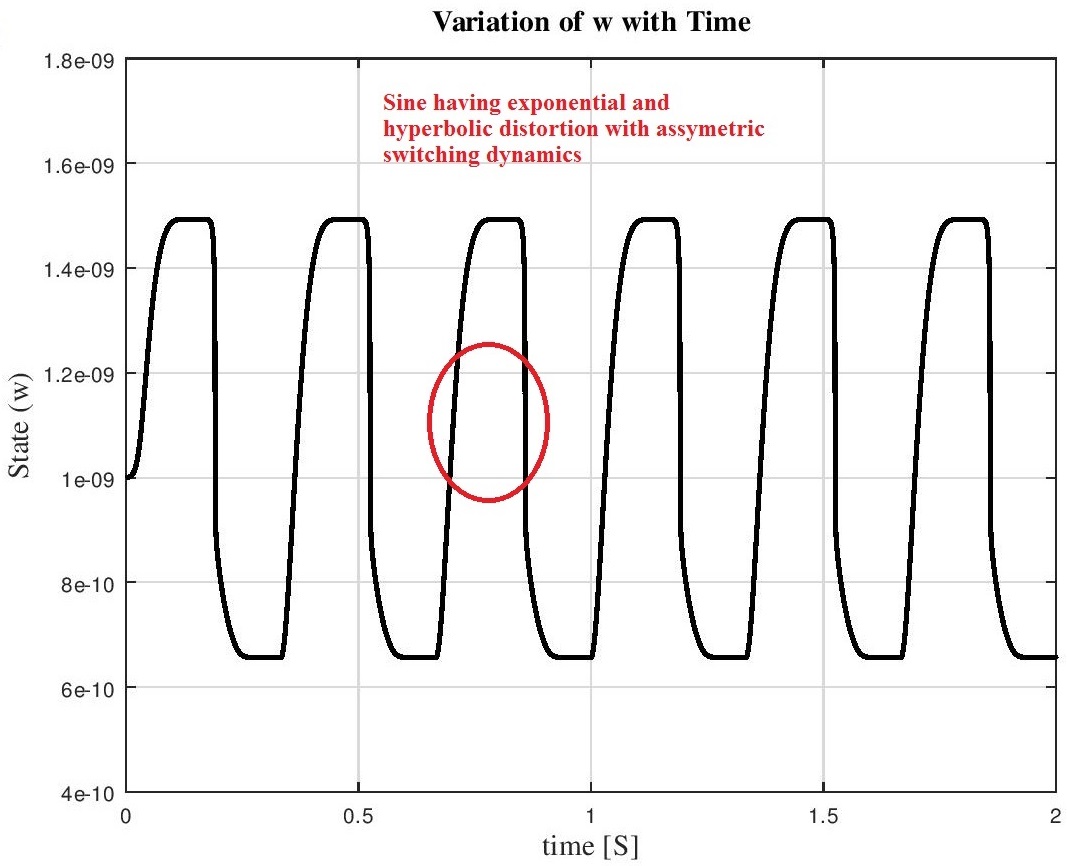}  
  \caption{Switching dynamics in  Pickett \textit{et al.} model.}
  \label{fig:6c}
\end{subfigure}
 \caption{Switching dynamics in three different memristor models.}
 \label{Fig: switchingdynamics}
\end{figure}

\begin{figure}[t]
\center
\includegraphics[width=\linewidth]{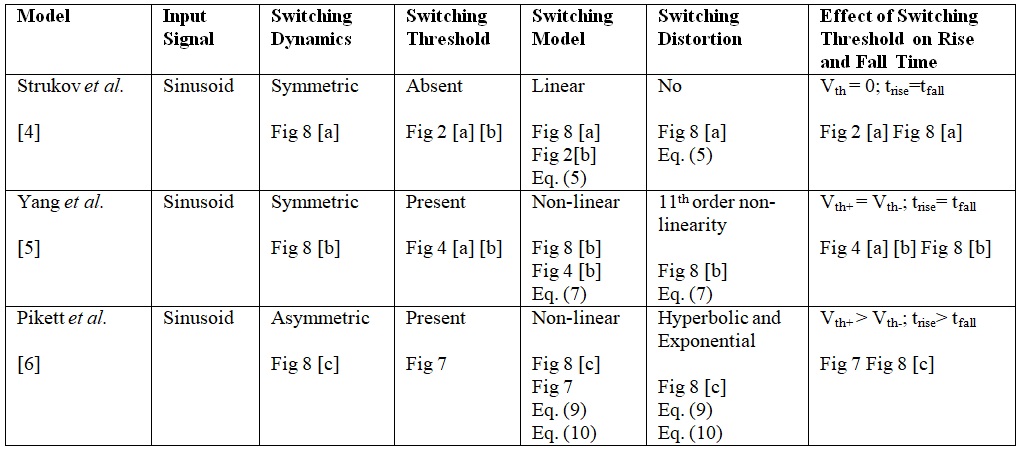}
\caption{Summary Table}
\label{Fig: Summary Table}
\end{figure}

We perform simulation on the memristor models proposed by \cite{stanley}, \cite{yang}, and \cite{pickett} to see the variation of state with time. 
The input is a sinusoidal signal with peak to peak amplitude enough to drive the memristors into saturation region (fully ON and OFF). The device dimension is taken as 10 $nm$ and resistance transfer ratio (ratio of $R_{ON}/R_{OFF} $) is 160 in all three models. The results are displayed in Fig. \ref{Fig: switchingdynamics}, and summarized in Fig. \ref{Fig: Summary Table}.

\section{Conclusion and Recommendations}
Strukov model shows linear and symmetric  switching dynamics and switching model respectively. 
Due to linear nature of this model, it could be useful in analog applications, which demand linear response.  
In Yang model, symmetric switching dynamics is seen, but non linearity in switching model can be controlled.
Since the non-linearity of Yang model is controllable, it could be useful in mixed signal type applications.
However, Pickett model has non linear and asymmetric  switching dynamics and switching model respectively. 
Due to non-linear nature of Pickett model, it is useful in logic circuit applications.

\end{document}